\documentclass[10pt,a4paper,onecolumn]{article}
\usepackage{marginnote}
\usepackage{graphicx}
\usepackage{xcolor}
\usepackage{authblk,etoolbox}
\usepackage{titlesec}
\usepackage{calc}
\usepackage{tikz}
\usepackage{hyperref}
\hypersetup{colorlinks,breaklinks=true,
            urlcolor=[rgb]{0.0, 0.5, 1.0},
            linkcolor=[rgb]{0.0, 0.5, 1.0}}
\usepackage{caption}
\usepackage{tcolorbox}
\usepackage{amssymb,amsmath}
\usepackage{ifxetex,ifluatex}
\usepackage{seqsplit}
\usepackage{xstring}

\usepackage{float}
\let\origfigure\figure
\let\endorigfigure\endfigure
\renewenvironment{figure}[1][2] {
    \expandafter\origfigure\expandafter[H]
} {
    \endorigfigure
}

\usepackage{fixltx2e} 
\usepackage[
  backend=biber,
]{biblatex}
\bibliography{paper.bib}

\let\textttOrig=\texttt
\def\texttt#1{\expandafter\textttOrig{\seqsplit{#1}}}
\renewcommand{\seqinsert}{\ifmmode
  \allowbreak
  \else\penalty6000\hspace{0pt plus 0.02em}\fi}

\makeatletter
\let\href@Orig=\href
\def\href@Urllike#1#2{\href@Orig{#1}{\begingroup
    \def\Url@String{#2}\Url@FormatString
    \endgroup}}
\def\href@Notdoi#1#2{\def\tempa{#1}\def\tempb{#2}%
  \ifx\tempa\tempb\relax\href@Urllike{#1}{#2}\else
  \href@Orig{#1}{#2}\fi}
\def\href#1#2{%
  \IfBeginWith{#1}{https://doi.org}%
  {\href@Urllike{#1}{#2}}{\href@Notdoi{#1}{#2}}}
\makeatother

\newlength{\cslhangindent}
\newlength{\csllabelwidth}
\newenvironment{CSLReferences}[3] 
 {
  \setlength{\parindent}{0pt}
  \ifodd #1 \everypar{\setlength{\hangindent}{\cslhangindent}}\ignorespaces\fi
  \ifnum #2 > 0
  \setlength{\parskip}{#2\baselineskip}
  \fi
 }%
 {}
\usepackage{calc}

\usepackage[top=3.5cm, bottom=3cm, right=1.5cm, left=1.0cm,
            headheight=2.2cm, reversemp, includemp, marginparwidth=4.5cm]{geometry}

\titleformat{\section}
  {\normalfont\sffamily\Large\bfseries}
  {}{0pt}{}
\titleformat{\subsection}
  {\normalfont\sffamily\large\bfseries}
  {}{0pt}{}
\titleformat{\subsubsection}
  {\normalfont\sffamily\bfseries}
  {}{0pt}{}
\titleformat*{\paragraph}
  {\sffamily\normalsize}

\usepackage{fancyhdr}
\makeatletter
\let\ps@plain\ps@fancy
\fancyheadoffset[L]{4.5cm}
\fancyfootoffset[L]{4.5cm}

\definecolor{linky}{rgb}{0.0, 0.5, 1.0}

\newtcolorbox{repobox}
   {colback=red, colframe=red!75!black,
     boxrule=0.5pt, arc=2pt, left=6pt, right=6pt, top=3pt, bottom=3pt}

\newcommand{\ExternalLink}{%
   \tikz[x=1.2ex, y=1.2ex, baseline=-0.05ex]{%
       \begin{scope}[x=1ex, y=1ex]
           \clip (-0.1,-0.1)
               --++ (-0, 1.2)
               --++ (0.6, 0)
               --++ (0, -0.6)
               --++ (0.6, 0)
               --++ (0, -1);
           \path[draw,
               line width = 0.5,
               rounded corners=0.5]
               (0,0) rectangle (1,1);
       \end{scope}
       \path[draw, line width = 0.5] (0.5, 0.5)
           -- (1, 1);
       \path[draw, line width = 0.5] (0.6, 1)
           -- (1, 1) -- (1, 0.6);
       }
   }

\patchcmd{\@maketitle}{center}{flushleft}{}{}
\patchcmd{\@maketitle}{center}{flushleft}{}{}
\patchcmd{\@maketitle}{\LARGE}{\LARGE\sffamily}{}{}
\def\maketitle{{%
  
  \AB@maketitle}}
\makeatletter
\renewcommand\AB@affilsepx{ \protect\Affilfont}
\renewcommand\AB@affilnote[1]{{\bfseries #1}\hspace{3pt}}
\renewcommand{\affil}[2][]%
   {\newaffiltrue\let\AB@blk@and\AB@pand
      \if\relax#1\relax\def\AB@note{\AB@thenote}\else\def\AB@note{#1}%
        \setcounter{Maxaffil}{0}\fi
        \begingroup
        \let\href=\href@Orig
        \let\texttt=\textttOrig
        \let\protect\@unexpandable@protect
        \def\thanks{\protect\thanks}\def\footnote{\protect\footnote}%
        \@temptokena=\expandafter{\AB@authors}%
        {\def\\{\protect\\\protect\Affilfont}\xdef\AB@temp{#2}}%
         \xdef\AB@authors{\the\@temptokena\AB@las\AB@au@str
         \protect\\[\affilsep]\protect\Affilfont\AB@temp}%
         \gdef\AB@las{}\gdef\AB@au@str{}%
        {\def\\{, \ignorespaces}\xdef\AB@temp{#2}}%
        \@temptokena=\expandafter{\AB@affillist}%
        \xdef\AB@affillist{\the\@temptokena \AB@affilsep
          \AB@affilnote{\AB@note}\protect\Affilfont\AB@temp}%
      \endgroup
       \let\AB@affilsep\AB@affilsepx
}
\makeatother

\renewcommand\Affilfont{\sffamily\small\mdseries}
\ifnum 0\ifxetex 1\fi\ifluatex 1\fi=0 
  \usepackage[T1]{fontenc}
  \usepackage[utf8]{inputenc}

\else 
  \ifxetex
    \usepackage{mathspec}
    \usepackage{fontspec}

  \else
    \usepackage{fontspec}
  \fi
  \defaultfontfeatures{Ligatures=TeX,Scale=MatchLowercase}

\fi
\IfFileExists{upquote.sty}{\usepackage{upquote}}{}
\IfFileExists{microtype.sty}{%
\usepackage{microtype}
\UseMicrotypeSet[protrusion]{basicmath} 
}{}

\usepackage{hyperref}
\hypersetup{unicode=true,
            pdftitle={GWPopulation: Hardware agnostic population inference for compact binaries and beyond},
            pdfborder={0 0 0},
            breaklinks=true}
\let\addcontentslineOrig=\addcontentsline
\def\addcontentsline#1#2#3{\bgroup
  \let\texttt=\textttOrig\addcontentslineOrig{#1}{#2}{#3}\egroup}
\let\markbothOrig\markboth
\def\markboth#1#2{\bgroup
  \let\texttt=\textttOrig\markbothOrig{#1}{#2}\egroup}
\let\markrightOrig\markright
\def\markright#1{\bgroup
  \let\texttt=\textttOrig\markrightOrig{#1}\egroup}

\IfFileExists{parskip.sty}{%
\usepackage{parskip}
}{
\setlength{\parindent}{0pt}
\setlength{\parskip}{6pt plus 2pt minus 1pt}
}
\providecommand{\tightlist}{%
  \setlength{\itemsep}{0pt}\setlength{\parskip}{0pt}}
\ifx\paragraph\undefined\else
\let\oldparagraph\paragraph
\renewcommand{\paragraph}[1]{\oldparagraph{#1}\mbox{}}
\fi
\ifx\subparagraph\undefined\else
\let\oldsubparagraph\subparagraph
\renewcommand{\subparagraph}[1]{\oldsubparagraph{#1}\mbox{}}
\fi

\title{GWPopulation: Hardware agnostic population inference for compact
binaries and beyond}

        \author[1]{Colm Talbot}
          \author[2]{Amanda Farah}
          \author[3, 4]{Shanika Galaudage}
          \author[5, 6]{Jacob Golomb}
          \author[7, 8]{Hui Tong}
    
      \affil[1]{Kavli Institute for Cosmological Physics, University of
Chicago, USA}
      \affil[2]{Department of Physics, University of Chicago, Chicago,
USA}
      \affil[3]{Laboratoire Lagrange, Université Côte d'Azur,
Observatoire de la Côte d'Azur, CNRS, Bd de l'Observatoire, 06300,
France}
      \affil[4]{Laboratoire Artemis, Université Côte d'Azur,
Observatoire de la Côte d'Azur, CNRS, Bd de l'Observatoire, 06300,
France}
      \affil[5]{Department of Physics, California Institute of
Technology, Pasadena, CA}
      \affil[6]{LIGO Laboratory, California Institute of Technology,
Pasadena, CA}
      \affil[7]{School of Physics and Astronomy, Monash University, VIC
3800, Australia}
      \affil[8]{OzGrav: The ARC Centre of Excellence for Gravitational
Wave Discovery, Clayton VIC 3800, Australia}
  \date{\vspace{-7ex}}

\begin{document}
\maketitle

\marginpar{

  \begin{flushleft}
  \sffamily\small

  {\bfseries DOI:} \href{https://doi.org/DOI unavailable}{\color{linky}{DOI unavailable}}

  \vspace{2mm}

  {\bfseries Software}
  \begin{itemize}
    \setlength\itemsep{0em}
    \item \href{N/A}{\color{linky}{Review}} \ExternalLink
    \item \href{NO_REPOSITORY}{\color{linky}{Repository}} \ExternalLink
    \item \href{DOI unavailable}{\color{linky}{Archive}} \ExternalLink
  \end{itemize}

  \vspace{2mm}

  \par\noindent\hrulefill\par

  \vspace{2mm}

  {\bfseries Editor:} \href{https://example.com}{Pending
Editor} \ExternalLink \\
  \vspace{1mm}
    {\bfseries Reviewers:}
  \begin{itemize}
  \setlength\itemsep{0em}
    \item \href{https://github.com/Pending Reviewers}{@Pending
Reviewers}
    \end{itemize}
    \vspace{2mm}

  {\bfseries Submitted:} N/A\\
  {\bfseries Published:} N/A

  \vspace{2mm}
  {\bfseries License}\\
  Authors of papers retain copyright and release the work under a Creative Commons Attribution 4.0 International License (\href{http://creativecommons.org/licenses/by/4.0/}{\color{linky}{CC BY 4.0}}).

  \end{flushleft}
}

\hypertarget{summary}{%
\section{Summary}\label{summary}}

Since the first direct detection of gravitational waves by the
LIGO--Virgo collaboration in 2015 (B. P. Abbott \& et al., 2016), the
size of the gravitational-wave transient catalog has grown to nearly 100
events (R. Abbott \& et al., 2023b), with more than as many observed
during the ongoing fourth observing run. Extracting
astrophysical/cosmological information from these observations is a
hierarchical Bayesian inference problem. \texttt{GWPopulation} is
designed to provide simple-to-use, robust, and extensible tools for
hierarchical inference in gravitational-wave astronomy/cosmology. It has
been widely adopted for gravitational-wave astrnomy, including producing
flagship results for the LIGO-Virgo-KAGRA collaborations (e.g., R.
Abbott \& et al. (2023a), The LIGO Scientific Collaboration et al.
(2024))\footnote{For a full listing of papers using
  \texttt{GWPopulation}, see the
  \href{https://ui.adsabs.harvard.edu/abs/2019PhRvD.100d3030T/citations}{citations
  for the previous publication}.}. While designed to work with
observations of compact binary coalescences, \texttt{GWPopulation} may
be available to a wider range of hierarchical Bayesian inference
problems.

Building on \texttt{Bilby} (Ashton et al., 2019), \texttt{GWPopulation}
can easily be used with a range of stochastic samplers through a
standard interface. By providing access to a range of array backends
(\texttt{numpy} (Harris et al., 2020), \texttt{JAX} (Bradbury et al.,
2018), and \texttt{cupy} (Okuta et al., 2017) are currently supported)
\texttt{GWPopulation} is hardware agnostic and can leverage hardware
acceleration to meet the growing computational needs of these analyses.
Included in the package are:

\begin{itemize}
\tightlist
\item
  implementations of the most commonly used likelihood functions in the
  field.
\item
  commonly used models for describing the astrophysical population of
  merging compact binaries. Including the ``PowerLaw+Peak'' and
  ``PowerLaw+Spline'' mass models, ``Default'' spin model, and
  ``PowerLaw'' redshift models used in the latest LIGO-Virgo-KAGRA
  collaboration analysis of the compact binary population.
\item
  functionality to simultaneously infer the astrophysical distribution
  of sources and cosmic expansion history using the ``spectral siren''
  method (Ezquiaga \& Holz, 2022).
\item
  a standard specification allowing users to define additional models.
\end{itemize}

\hypertarget{statement-of-need}{%
\section{Statement of need}\label{statement-of-need}}

Hierarchical Bayesian inference is the standard method for inferring
parameters describing the astrophysical population of compact binaries
and the cosmic expansion history (e.g., Thrane \& Talbot (2019), Vitale
et al. (2022)). The first step in the hierarchical inference process is
drawing samples from the posterior distributions for the source
parameters of each event under a fiducial prior distribution along with
a set of simulated signals used to quantify the sensitivity of
gravitational-wave searches. Next, these samples are used to estimate
the population likelihood using Monte Carlo integration with a
computational cost that grows quadratically with the size of the
observed population. Since evaluating these Monte Carlo integrals is
embarrassingly parallel, this is a prime candidate for hardware
acceleration using graphics/tensor processing units.
\texttt{GWPopulation} provides functionality need to perform this second
step.

Maximizing the information we can extract from the gravitational-wave
transient catalog requires a framework where potential population models
can be quickly constrained with the observed data with minimal
boilerplate code. Additionally, the availability of a standard
open-source implementation improves the reliability and reproducibility
of published results. \texttt{GWPopulation} addresses all of these
points by providing a standard, open-source, implementation of the
standard functionality needed to perform population analyses while
enabling user-defined models to be provided by a \texttt{Python}
function/class definition. The flexible backend system means hardware
acceleration can be used with minimal coding effort. Using
\texttt{GWPopulation} on Google Colab, it is possible to perform an
exploratory analysis with a new population model in minutes and produce
production-quality results without needing high-performance/throughput
computing clusters. With access to high throughput computing resources,
a wide range of potential models can be easily explored using the
associated \texttt{gwpopulation\_pipe} (Talbot, 2021) package.

\hypertarget{related-packages}{%
\section{Related packages}\label{related-packages}}

Several other packages are actively used and maintained in the community
that can be used for population inference that operate in complementary
ways to \texttt{GWPopulation}.

\begin{itemize}
\tightlist
\item
  \texttt{GWInferno} (Edelman et al., 2023) is a package for
  hierarchical inference with gravitational-wave sources intended for
  use with \texttt{numpyro} targeting high-dimensional models.
  \texttt{GWInferno} includes many population models initially adapted
  from \texttt{GWPopulation}.
\item
  There are a wide range of packages designed for joint astrophyical and
  cosmological inference with gravitational-wave transients including
  \texttt{icarogw} (Mastrogiovanni et al., 2024), \texttt{gwcosmo} (Gray
  et al., 2023), \texttt{MGCosmoPop} (Mancarella \& Genoud-Prachex,
  2022), and \texttt{CHIMERA} (Borghi et al., 2024). \texttt{icarogw}
  supports some harware acceleration using \texttt{cupy} but some
  cosmological calculations are limited to CPU support only.
  \texttt{chimera} is \texttt{JAX}-compatible and supports flat
  Lambda-CDM cosmologies along with analysis using galaxy catalogs.
\item
  \texttt{vamana} (Tiwari, 2021) models the compact binary distribution
  as a mixture of Gaussians and power-law distributions,
  \texttt{popmodels} (Wysocki \& O'Shaughnessy, 2017--) implements a
  range of parametric models for the compact binary distribution and
  supports sampling via \texttt{emcee} (Foreman-Mackey et al., 2013),
  neither supports hardware acceleration at the time of writing.
\end{itemize}

\hypertarget{acknowledgements}{%
\section{Acknowledgements}\label{acknowledgements}}

CT is supported by an Eric and Wendy Schmidt AI in Science Fellowship, a
Schmidt Sciences program. SG is supported by the ANR COSMERGE project,
grant ANR-20-CE31-001 of the French Agence Nationale de la Recherche. AF
is supported by the NSF Research Traineeship program under grant
No.~DGE1735359, and by the National Science Foundation Graduate Research
Fellowship Program under Grant No.~DGE-1746045. HT is supported by
Australian Research Council (ARC) Centre of Excellence CE230100016.

\hypertarget{references}{%
\section*{References}\label{references}}
\addcontentsline{toc}{section}{References}

\hypertarget{refs}{}
\begin{CSLReferences}{1}{0}
\leavevmode\hypertarget{ref-GW150914}{}%
Abbott, B. P., \& et al. (2016). Observation of gravitational waves from
a binary black hole merger. \emph{Phys. Rev. Lett.}, \emph{116}, 061102.
\url{https://doi.org/10.1103/PhysRevLett.116.061102}

\leavevmode\hypertarget{ref-GWTC3Pop}{}%
Abbott, R., \& et al. (2023a). Population of merging compact binaries
inferred using gravitational waves through GWTC-3. \emph{Phys. Rev. X},
\emph{13}, 011048. \url{https://doi.org/10.1103/PhysRevX.13.011048}

\leavevmode\hypertarget{ref-GWTC3}{}%
Abbott, R., \& et al. (2023b). {GWTC-3: Compact Binary Coalescences
Observed by LIGO and Virgo during the Second Part of the Third Observing
Run}. \emph{Physical Review X}, \emph{13}(4), 041039.
\url{https://doi.org/10.1103/PhysRevX.13.041039}

\leavevmode\hypertarget{ref-Bilby}{}%
Ashton, G., Hübner, M., Lasky, P. D., Talbot, C., \& et al. (2019).
{BILBY: A User-friendly Bayesian Inference Library for
Gravitational-wave Astronomy}. \emph{241}(2), 27.
\url{https://doi.org/10.3847/1538-4365/ab06fc}

\leavevmode\hypertarget{ref-chimera}{}%
Borghi, N., Mancarella, M., Moresco, M., Tagliazucchi, M., Iacovelli,
F., Cimatti, A., \& Maggiore, M. (2024). {Cosmology and Astrophysics
with Standard Sirens and Galaxy Catalogs in View of Future Gravitational
Wave Observations}. \emph{964}(2), 191.
\url{https://doi.org/10.3847/1538-4357/ad20eb}

\leavevmode\hypertarget{ref-jax}{}%
Bradbury, J., Frostig, R., Hawkins, P., Johnson, M. J., Leary, C.,
Maclaurin, D., Necula, G., Paszke, A., VanderPlas, J., Wanderman-Milne,
S., \& Zhang, Q. (2018). \emph{{JAX}: Composable transformations of
{P}ython+{N}um{P}y programs} (Version 0.3.13) {[}Computer software{]}.
\url{http://github.com/google/jax}

\leavevmode\hypertarget{ref-gwinferno}{}%
Edelman, B., Farr, B., \& Doctor, Z. (2023). {Cover Your Basis:
Comprehensive Data-driven Characterization of the Binary Black Hole
Population}. \emph{946}(1), 16.
\url{https://doi.org/10.3847/1538-4357/acb5ed}

\leavevmode\hypertarget{ref-Ezquiaga2022}{}%
Ezquiaga, J. M., \& Holz, D. E. (2022). {Spectral Sirens: Cosmology from
the Full Mass Distribution of Compact Binaries}. \emph{Phys. Rev.
Lett.}, \emph{129}(6), 061102.
\url{https://doi.org/10.1103/PhysRevLett.129.061102}

\leavevmode\hypertarget{ref-emcee}{}%
Foreman-Mackey, D., Hogg, D. W., Lang, D., \& Goodman, J. (2013). Emcee:
The MCMC hammer. \emph{PASP}, \emph{125}, 306--312.
\url{https://doi.org/10.1086/670067}

\leavevmode\hypertarget{ref-gwcosmo}{}%
Gray, R., Beirnaert, F., Karathanasis, C., Revenu, B., Turski, C., Chen,
A., Baker, T., Vallejo, S., Romano, A. E., Ghosh, T., Ghosh, A., Leyde,
K., Mastrogiovanni, S., \& More, S. (2023). Joint cosmological and
gravitational-wave population inference using dark sirens and galaxy
catalogues. \emph{Journal of Cosmology and Astroparticle Physics},
\emph{2023}(12), 023.
\url{https://doi.org/10.1088/1475-7516/2023/12/023}

\leavevmode\hypertarget{ref-numpy}{}%
Harris, C. R., Millman, K. J., Walt, S. J. van der, Gommers, R.,
Virtanen, P., Cournapeau, D., Wieser, E., Taylor, J., Berg, S., Smith,
N. J., Kern, R., Picus, M., Hoyer, S., Kerkwijk, M. H. van, Brett, M.,
Haldane, A., Río, J. F. del, Wiebe, M., Peterson, P., \ldots{} Oliphant,
T. E. (2020). Array programming with {NumPy}. \emph{Nature},
\emph{585}(7825), 357--362.
\url{https://doi.org/10.1038/s41586-020-2649-2}

\leavevmode\hypertarget{ref-mgcosmo}{}%
Mancarella, M., \& Genoud-Prachex, E. (2022).
\emph{CosmoStatGW/MGCosmoPop: v1.0.0} (Version v1.0.0) {[}Computer
software{]}. Zenodo. \url{https://doi.org/10.5281/zenodo.6323173}

\leavevmode\hypertarget{ref-icarogw}{}%
Mastrogiovanni, S., Pierra, G., Perriès, S., Laghi, D., Santoro, G. C.,
Ghosh, A., Gray, R., Karathanasis, C., \& Leyde, K. (2024). {ICAROGW: A
python package for inference of astrophysical population properties of
noisy, heterogeneous, and incomplete observations}. \emph{682}, A167.
\url{https://doi.org/10.1051/0004-6361/202347007}

\leavevmode\hypertarget{ref-cupy}{}%
Okuta, R., Unno, Y., Nishino, D., Hido, S., \& Loomis, C. (2017). CuPy:
A NumPy-compatible library for NVIDIA GPU calculations.
\emph{Proceedings of Workshop on Machine Learning Systems (LearningSys)
in the Thirty-First Annual Conference on Neural Information Processing
Systems (NIPS)}.
\url{http://learningsys.org/nips17/assets/papers/paper_16.pdf}

\leavevmode\hypertarget{ref-gwpop_pipe}{}%
Talbot, C. (2021). \emph{GWPopulation pipe}.
\url{https://doi.org/10.5281/zenodo.5654673}

\leavevmode\hypertarget{ref-GW230529}{}%
The LIGO Scientific Collaboration, the Virgo Collaboration, \& the KAGRA
Collaboration. (2024). {Observation of Gravitational Waves from the
Coalescence of a \(2.5-4.5~M_\odot\) Compact Object and a Neutron Star}.
\emph{arXiv e-Prints}, arXiv:2404.04248.
\url{https://doi.org/10.48550/arXiv.2404.04248}

\leavevmode\hypertarget{ref-Thrane2019}{}%
Thrane, E., \& Talbot, C. (2019). {An introduction to Bayesian inference
in gravitational-wave astronomy: Parameter estimation, model selection,
and hierarchical models}. \emph{36}, e010.
\url{https://doi.org/10.1017/pasa.2019.2}

\leavevmode\hypertarget{ref-vamana}{}%
Tiwari, V. (2021). {VAMANA: modeling binary black hole population with
minimal assumptions}. \emph{Classical and Quantum Gravity},
\emph{38}(15), 155007. \url{https://doi.org/10.1088/1361-6382/ac0b54}

\leavevmode\hypertarget{ref-Vitale2022}{}%
Vitale, S., Gerosa, D., Farr, W. M., \& Taylor, S. R. (2022). {Inferring
the Properties of a Population of Compact Binaries in Presence of
Selection Effects}. In \emph{Handbook of gravitational wave astronomy}
(p. 45). \url{https://doi.org/10.1007/978-981-15-4702-7_45-1}

\leavevmode\hypertarget{ref-popmodels}{}%
Wysocki, D., \& O'Shaughnessy, R. (2017--). \emph{Bayesian parametric
population models}.
\href{https://bayesian-parametric-population-models.readthedocs.io}{bayesian-parametric-population-models.readthedocs.io}

\end{CSLReferences}

\end{document}